\documentclass[preprint,aps,superscriptaddress]{revtex4}

\usepackage{graphicx}
\usepackage{epsfig}
\usepackage{dcolumn}
\usepackage{bm}
\usepackage{natbib}
\usepackage{color}

\begin{document}

\title{Anisotropic nuclear-spin diffusion in double quantum wells}

\author{T. Hatano}
\email{hatano@m.tohoku.ac.jp}
\affiliation{Department of Physics, Tohoku University, Sendai 980-8578, Japan}
\affiliation{JST, ERATO, Nuclear Spin Electronics Project, Sendai 980-8578, Japan}

\author{W. Kume}
\altaffiliation[Present address:]{Tokyo Electron America Inc.}
\affiliation{Department of Physics, Tohoku University, Sendai 980-8578, Japan}

\author{S. Watanabe}
\altaffiliation[Present address:]{Bio-AFM Frontier Research Center, Kanazawa University}
\affiliation{JST, ERATO, Nuclear Spin Electronics Project, Sendai 980-8578, Japan}

\author{K. Akiba}
\altaffiliation[Present address:]{ Department of Applied Physics, Tokyo University of Agriculture and Technology}
\affiliation{Department of Physics, Tohoku University, Sendai 980-8578, Japan}
\affiliation{JST, ERATO, Nuclear Spin Electronics Project, Sendai 980-8578, Japan}

\author{K. Nagase}
\affiliation{Department of Physics, Tohoku University, Sendai 980-8578, Japan}
\affiliation{JST, ERATO, Nuclear Spin Electronics Project, Sendai 980-8578, Japan}

\author{Y. Hirayama}
\email{hirayama@m.tohoku.ac.jp}
\affiliation{Department of Physics, Tohoku University, Sendai 980-8578, Japan}
\affiliation{JST, ERATO, Nuclear Spin Electronics Project, Sendai 980-8578, Japan}
\affiliation{WPI-AIMR, Tohoku University, Sendai 980- 8577, Japan}

\date{\today}

\begin{abstract}
Nuclear-spin diffusion in double quantum wells (QWs) is examined by using dynamic nuclear polarization (DNP) at a Landau level filling factor $\nu=2/3$ spin phase transition (SPT). The longitudinal resistance increases during the DNP of one of the two QW (the polarization QW) by means of a large applied current and starts to decrease just after the termination of the DNP. On the other hand, the longitudinal resistance of the other QW (the detection QW) continuously increases for approximately 2h after the termination of the DNP of the polarization QW. It is therefore concluded that the nuclear spins diffuse from the polarization QW to the detection QW. The time evolution of the longitudinal resistance of the polarization QW is explained mainly by the nuclear-spin diffusion in the in-plane direction. In contrast, that of the detection QW manifests much slower nuclear diffusion in the perpendicular direction through the AlGaAs barrier.
\end{abstract}

\maketitle

\section{Introduction}

The spin degree of freedom has attracted enormous interest from the perspective of quantum computation based on nuclear spins \cite{Kane,Ladd,Pla} and electron spins \cite{Loss,Petta,Koppens,Brunner,Shulman}. Fluctuation of nuclear polarization is an important issue concerning nuclear-spin-based coherent systems. It also influences electron-spin-based systems because of hyperfine interaction between electron spins and nuclear spins. It is therefore important to elucidate the relaxation of nuclear spins, particularly, nuclear-spin diffusion caused by dipole-dipole interaction of nuclear spins in such solid-state systems. 

Nuclear-spin diffusion has been studied by using optical detection \cite{Paget,Malinowski,Nikolaenko}. It has also been studied electrically by using Hall bar devices \cite{Nakajima} and double quantum dots \cite{Reilly}. In these reports, nuclear-spin diffusion is discussed isotropically or in a one-dimensional direction only. However, in reality, nuclear spins are polarized locally, and nuclear-spin diffusion is expected to occur anisotropically in nanoscale devices such as quantum Hall devices \cite{Machida}, quantum point contacts \cite{Yusa,Ren}, and double quantum dots \cite{Ono} because of the existence of heterostructures and inhomogeneous electron distributions in the in-plane direction. However, anisotropic nuclear-spin diffusion has not been observed.

Recently, double quantum wells (QWs) have been used to measure nuclear-spin diffusion directly \cite{Tsuda,Hatano,Nguyen}. In double QWs, nuclear spins are polarized in one of the two QWs (namely, the gpolarization QWh), and the nuclear-spin polarization caused by diffusion from the polarization QW can be detected in the other QW (namely, the gdetection QWh). To polarize and detect nuclear spins in double QWs, hyperfine interaction at a Landau level filling factor $\nu=2/3$ spin phase transition (SPT) was used in the same manner as previous studies \cite{Kronmuller,Smet1,Smet2,Hashimoto}. By association with the SPT, an electron-spin polarized and unpolarized domain structure is formed \cite{Smet1,Smet2,Hashimoto,Kraus,Stern}, and dynamic nuclear polarization (DNP) is achieved by applying a large current at $\nu=2/3$ SPT, where the electron-nuclear spin flip-flop process occurs around the domain walls. Then, variation of the longitudinal resistances of the QWs is used to detect the nuclear-spin polarization \cite{Smet1,Smet2,Hashimoto,Kraus,Stern}. In such a manner, Nguyen {\it et al.} \cite{Nguyen} obtained a nuclear-spin-diffusion coefficient of about $15{\rm nm^2/s}$, which is consistent with the value measured optically for bulk GaAs \cite{Paget}. In their study, the nuclear polarization inside the QWs after DNP is assumed to be homogeneous. However, it is clear from our previous studies that the DNP attributed to the domain structure cannot cause the homogeneous polarization \cite{Hirayama}. In spite of the advantage of an inhomogeneous nuclear-spin distribution in a study of anisotropic nuclear-spin diffusion, this advantage was ignored in Ref.\cite{Nguyen}. Furthermore, in their experiment, it was difficult to avoid modifying the nuclear-spin diffusion by electrons existing in the QWs caused by an artifact arising from multi-measurement of the longitudinal resistance (see the Appendix A). It is thus important to avoid such an artifact in the measurement.

\section{Quantum Hall effect of two quantum wells}

We use a Hall bar device fabricated from a wafer with a GaAs/Al$_{0.33}$Ga$_{0.67}$As double-QW heterostructure. The widths of the two QWs and the potential barrier are 20nm and 2.2nm, respectively\cite{SAS}. The length (width) of the Hall bar is 100${\rm \mu m}$ ($30{\rm \mu m}$). The front and back gates were formed to tune the electron densities in the front and back QWs, respectively. Although the ohmic contacts are common to both QWs, it is possible to independently measure the current flowing through the back or front QWs by tuning two gate voltages. The mobility of the back (front) QW is approximately $150(180){\rm m^2/Vs}$ at electron density of $1\times 10^{15}\rm{m^{-2}}$. The Hall bar device was measured in a dilution refrigerator at a base temperature of approximately 30mK by lock-in detection at a frequency 13.4Hz, and a magnetic field $B$ was applied perpendicularly to the device. 

\begin{figure}[htp]
\includegraphics[width=1\columnwidth]{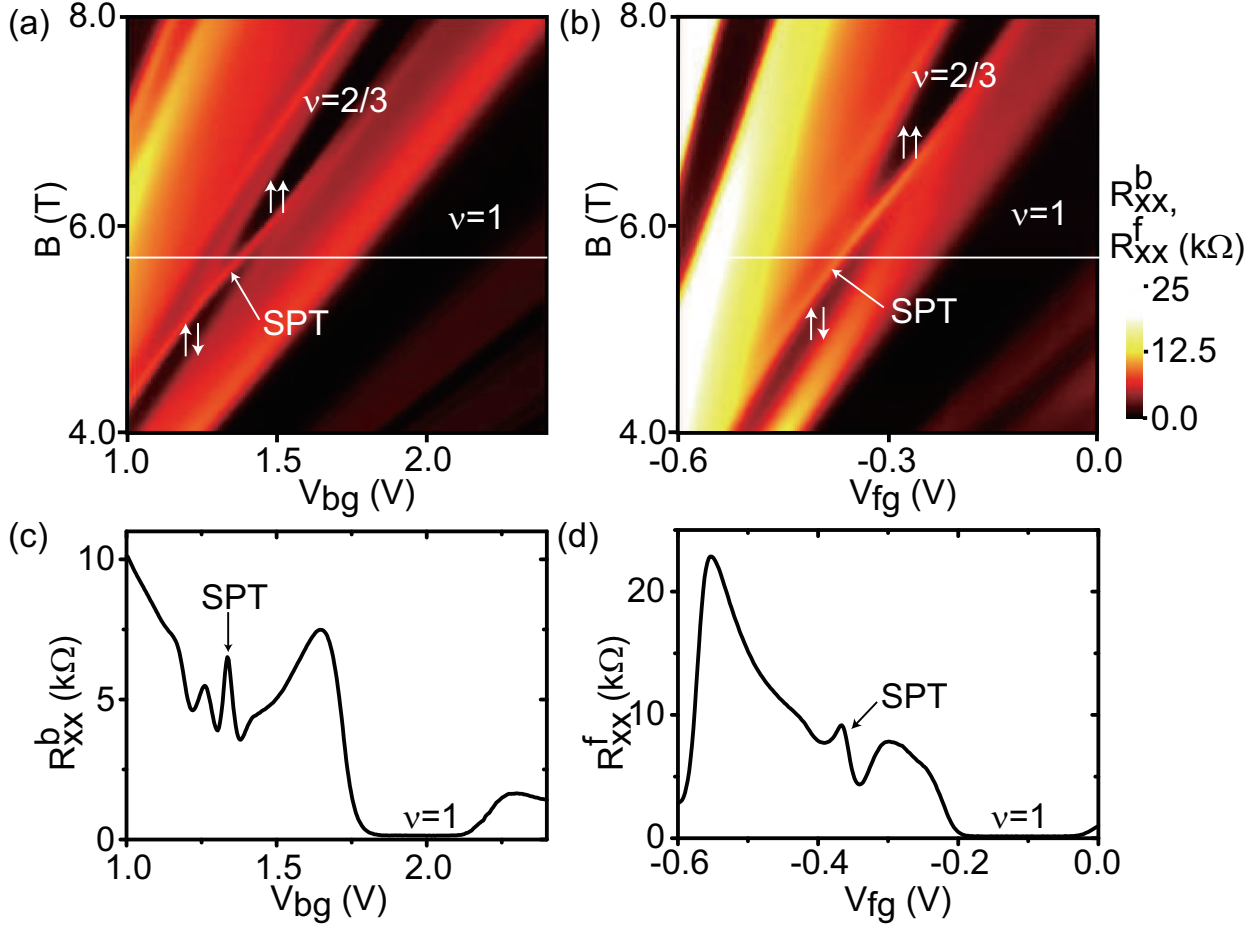}
\caption{(a) Longitudinal resistance of the back quantum well (QW), $R_{xx}^{b}$, as a function of back gate voltage $V_{bg}$ and magnetic field $B$ at front gate voltage $V_{fg}=-0.8$ V. The white line indicates $V_{bg}$ sweeping shown in (c). (b) Longitudinal resistance of the front QW, $R_{xx}^{f}$, as a function of $V_{fg}$ and $B$ at $V_{bg}=-1.2$ V. The white line indicates $V_{fg}$ sweeping shown in (d). (c) $R_{xx}^{b}$ as a function of $V_{bg}$ at $V_{fg}=-1.2$~V and $B=5.7$ T. (d) $R_{xx}^{f}$ as a function of $V_{fg}$ at $V_{bg}=-0.8$ V and $B=5.7$ T. 
} 
\label{QHE}
\end{figure}

Longitudinal resistance of the back (front) quantum well (QW),  $R_{xx}^{b}$ ($R_{xx}^{f}$), is shown as a function of the back (front) gate voltage $V_{bg}$ ($V_{fg}$) and a magnetic field $B$ at a current $I=10$ nA and the front (back) gate voltage $V_{fg}=-0.8$ V ($V_{bg}=-1.2$ V) in Figs.~\ref{QHE}(a)((b)). The black regions in Figs.~\ref{QHE}(a) and (b) indicate the QHE regions; particularly, the $\nu=1$ QHE is clearly observed. For the $\nu=2/3$ regions in Figs.~\ref{QHE} (a) and (b), the spin phase transition (SPT) peaks, which correspond to the boundary between the electron spin polarized and unpolarized regions \cite{Kronmuller}, can be seen. In Figs.~\ref{QHE}(a) and (b), the QHE of a single QW (without any effect arising from electrons in the other QW) can be seen. Therefore, the QHE of the back and front QWs is realized at $V_{fg}=-0.8$ V and $V_{bg}=-1.2$ V, respectively; that is, the front (back) QW is fully depleted at $V_{fg}=-0.8$ V ($V_{bg}=-1.2$ V). 

$R_{xx}^{b} (R_{xx}^{f})$ was measured as a function of $V_{bg}$ ($V_{fg}$) at $V_{fg}=-1.2$ V ($V_{bg}=-0.8$ V) and $B=5.7$ T for $I=1$ nA in Figs.~\ref{QHE}(c)((d)). Here the front (back) QW is fully depleted at $V_{fg}=-1.2$ V ($V_{bg}=-0.8$ V) in Figs.~\ref{QHE}(c)((d)). The SPT peak is clearly observed at $V_{bg}=1.326$ V ($V_{fg}=-0.373$ V) in Figs.~\ref{QHE}(c)((d)).

When a large current flows through the QW at the SPT in the $\nu=2/3$, the longitudinal resistance increases gradually because of the hyperfine interaction with nuclear spins; the nuclear spins in the QW are polarized \cite{Kronmuller,Smet1,Smet2,Hashimoto}. 
In contrast, the value of the longitudinal resistance is almost constant for a small current, which means negligible nuclear polarization. We confirmed that the nuclear spins were dynamically polarized at I = 50 nA and were not dynamically polarized at $I=1$ nA in both QWs (not shown). Therefore, we apply $I=50(1)$ nA to the QWs to polarize (detect) the nuclear spins.

\section{Measurement of nuclear-spin diffusion}

\begin{figure}[htb]
\includegraphics[width=1\columnwidth]{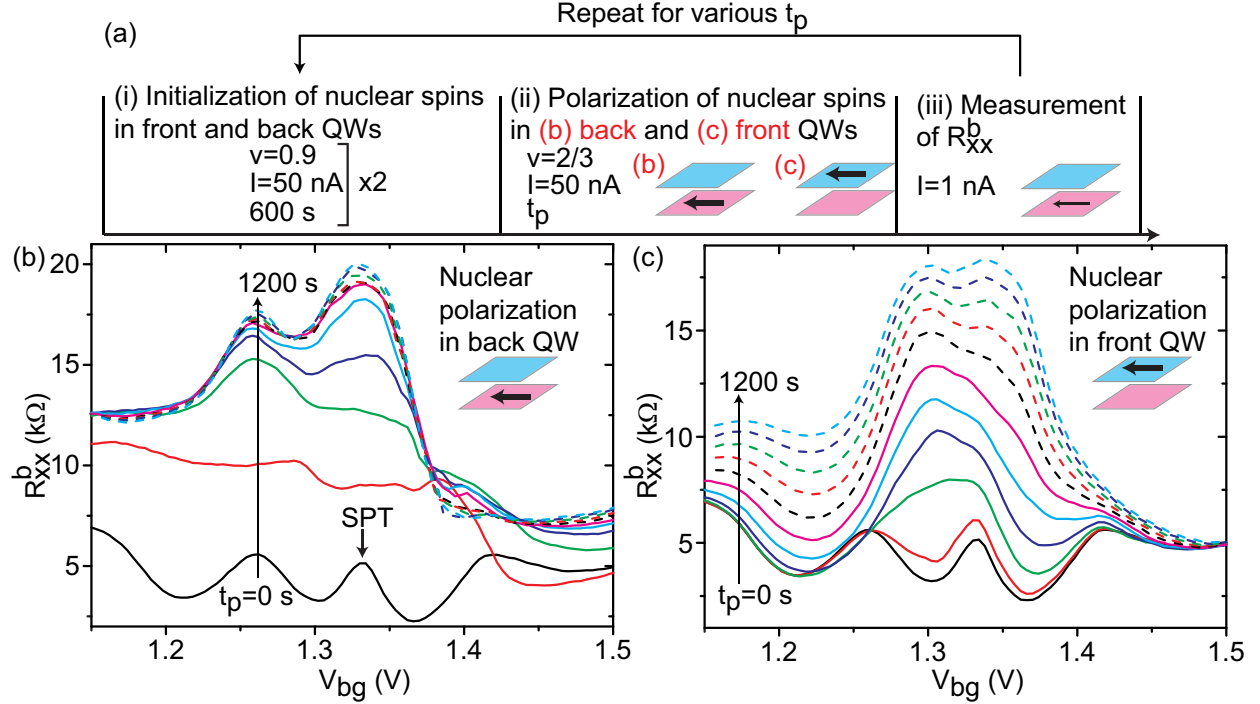}
\caption{
(a) Measurement sequence for time evolution of longitudinal resistance of the back QW, $R_{xx}^{b}$. Thick arrows indicate dynamic nuclear polarization under a current $I=50$nA, and the thin arrow indicates measurement under $I=1$nA. (b) $R_{xx}^{b}$ as a function of $V_{bg}$ after nuclear polarization in the back QW. The polarization time $t_p$ is changed from 0 to 1200s in steps of 120s. (c) Same as (b) except that the nuclear spins in the front QW is polarized.
} 
\label{figure2}
\end{figure}

To investigate how the polarized nuclear spins in the front and back QWs affect the longitudinal resistance of the back QW, $R_{xx}^{b}$, $R_{xx}^{b}$ is measured by following the measurement sequence shown in Fig.\ref{figure2}(a). First, the nuclear spins in the front and back QWs are initialized alternately; at $\nu=0.9$ and $I=50$nA applied to the front (back) QW for 600s, nuclear spins in the front (back) QWs are depolarized because of the Goldstone mode of the Skyrmions \cite{Smet2,Hashimoto}, and then the back (front) QW is depleted. Subsequently, the nuclear spins in the back (front) QW are polarized dynamically for polarization time $t_p$ by applying $I=50$nA at the SPT of the back (front) QW. The other QW is depleted during the nuclear polarization. Finally, $R_{xx}^{b}$ around the SPT is measured as a function of $V_{bg}$, at $I=1$nA depleting the front QW \cite{initial}. $R_{xx}^{b}$ after DNP in the back and front QWs for various values of $t_p$ is shown in Figs.\ref{figure2}(b) and (c), respectively. 
In Fig.\ref{figure2}(b), $R_{xx}^{b}$ around the SPT increases with $t_p$ and saturates at $t_p\gtrsim 500$s. This results from the current-induced DNP in the back QW \cite{Kronmuller,Smet1,Smet2,Hashimoto}. In Fig.\ref{figure2}(c), as $t_p$ increases, $R_{xx}^{b}$ around the SPT gradually increases in spite of no current flowing in the back QW. This $R_{xx}^{b}$ increase results from diffusion of polarized nuclear spins in the front QW to the back QW. The shape of $R_{xx}^{b}$ in (c) differs from that in (b), although $R_{xx}^{b}$ is measured in both cases. This result is discussed in more detail later.

\begin{figure}[htb]
\includegraphics[width=1\columnwidth]{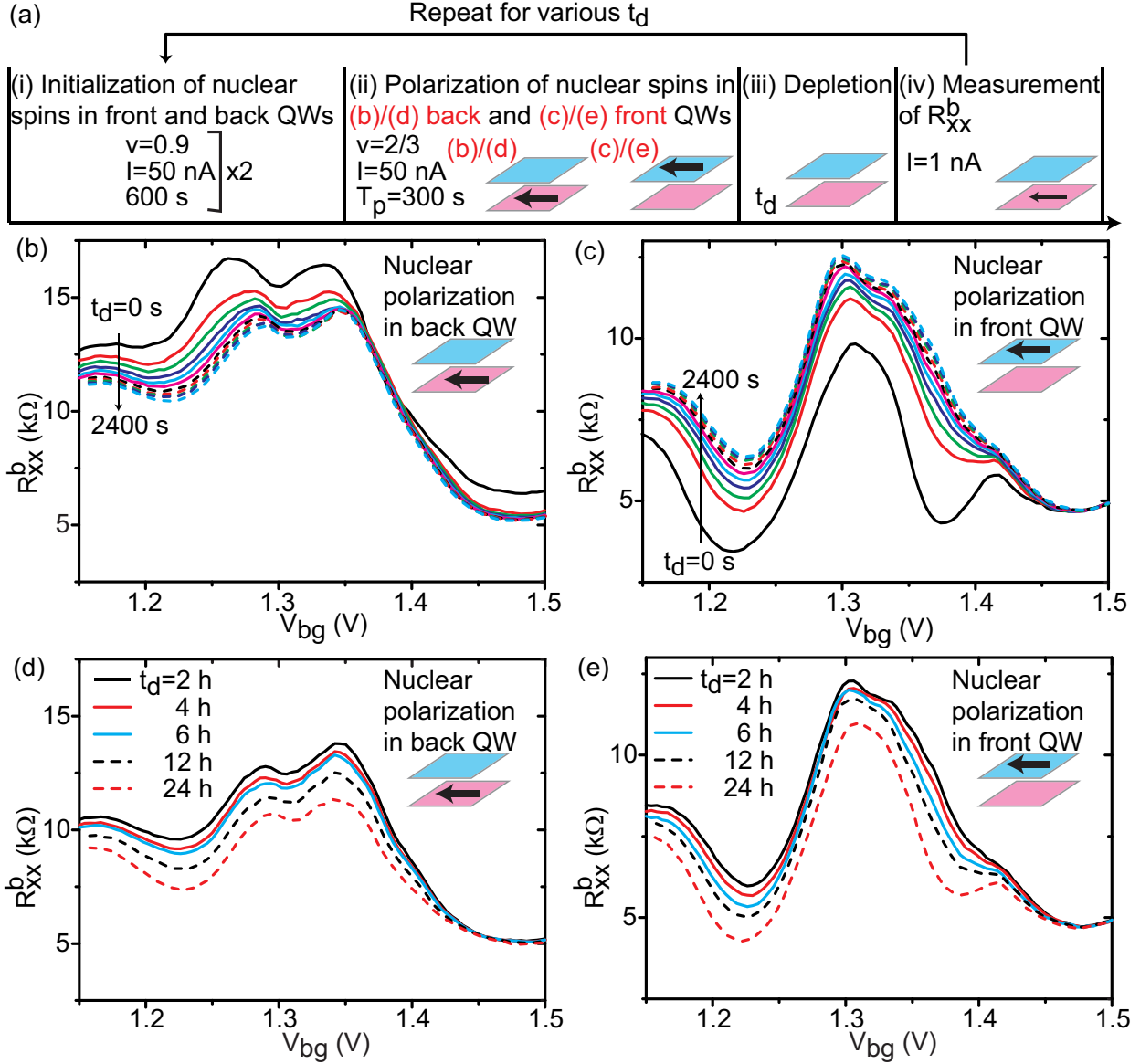}
\caption{(a) Measurement sequence for time evolution of $R_{xx}^{b}$. The thick and thin arrows mean the same as in Fig.\ref{figure2} (a). (b) $R_{xx}^{b}$ as a function of $V_{bg}$ for depletion time $t_d$ from 0 to 2400s in steps of 240s after nuclear polarization in the back QW. (c) Same as (b) except that the nuclear spins in the front QW is polarized. (d) $R_{xx}^{b}$ as a function of $V_{bg}$ for long $t_d$ after nuclear polarization in the back QW. (e) Same as (d) except that the nuclear spins in the front QW is polarized.
} 
\label{figure4}
\end{figure}

To clarify the nuclear-spin diffusion specifically, $R_{xx}^{b}$ was measured by following the measurement sequence shown in Fig.\ref{figure4}(a). First, the nuclear spins in the two QWs are initialized in the same manner as shown in Fig.\ref{figure2}(a). Subsequently, the nuclear spins in the back (front) QW are polarized for polarization time $T_p=300$s by applying $I=50$nA \cite{tp}. The other QW is depleted during the polarization. Next, both QWs are depleted for depletion time $t_d$. Finally, $R_{xx}^{b}$ around the SPT at $I=1$nA is measured depleting the front QW. $R_{xx}^{b}$ at various values of $t_d$, when the nuclear spins in the back QW are polarized, is shown in Fig.\ref{figure4}(b). $R_{xx}^{b}$ decreases around the SPT as $t_d$ increases, indicating nuclear-spin relaxation. In contrast, when the nuclear spins in the front QW are polarized, $R_{xx}^{b}$ at various values of $t_d$ (shown in Fig.\ref{figure4}(c)) increase in spite of a lack of direct current-induced DNP in the back QW. It clearly continues to increase even after nuclear polarization in the front QW is terminated. This result indicates the nuclear-spin diffusion from the front to the back QWs more specifically \cite{replace}. $R_{xx}^{b}$ at very long $t_d$ is shown in Figs.\ref{figure4} (d) and (e). As shown in Fig.\ref{figure4}(d), $R_{xx}^{b}$ continues to decrease. Although $R_{xx}^{b}$ continues to increase before 2400s (Fig.\ref{figure4}(c)), in Fig.\ref{figure4}(e), $R_{xx}^{b}$ turns to decrease after $\sim 2{\rm h}$. 
This is because the effect of the nuclear spin diffusion to the back QW becomes gradually weak after the termination of the DNP in the front QW. 
We will discuss, in more detail, the time evolution of $R_{xx}^{b}$ by using the data in Fig.~\ref{figure6}(c), which is calculated from the experimental results in Figs.~\ref{figure4} (b)-(e), later. 
In both cases, $R_{xx}^{b}$ does not decay fully, even at $t_d\sim 24{\rm h}$.

\section{Estimation of nuclear-spin diffusion coefficients}

To discuss the details of the experimentally obtained results, the nuclear-spin-diffusion coefficient in the double QW is estimated from our data. We consider the data shown in Fig.\ref{figure4}, in which case electrons in the double QWs are fully depleted during the nuclear-spin diffusion. 

\begin{figure}[htb]
\includegraphics[width=1\columnwidth]{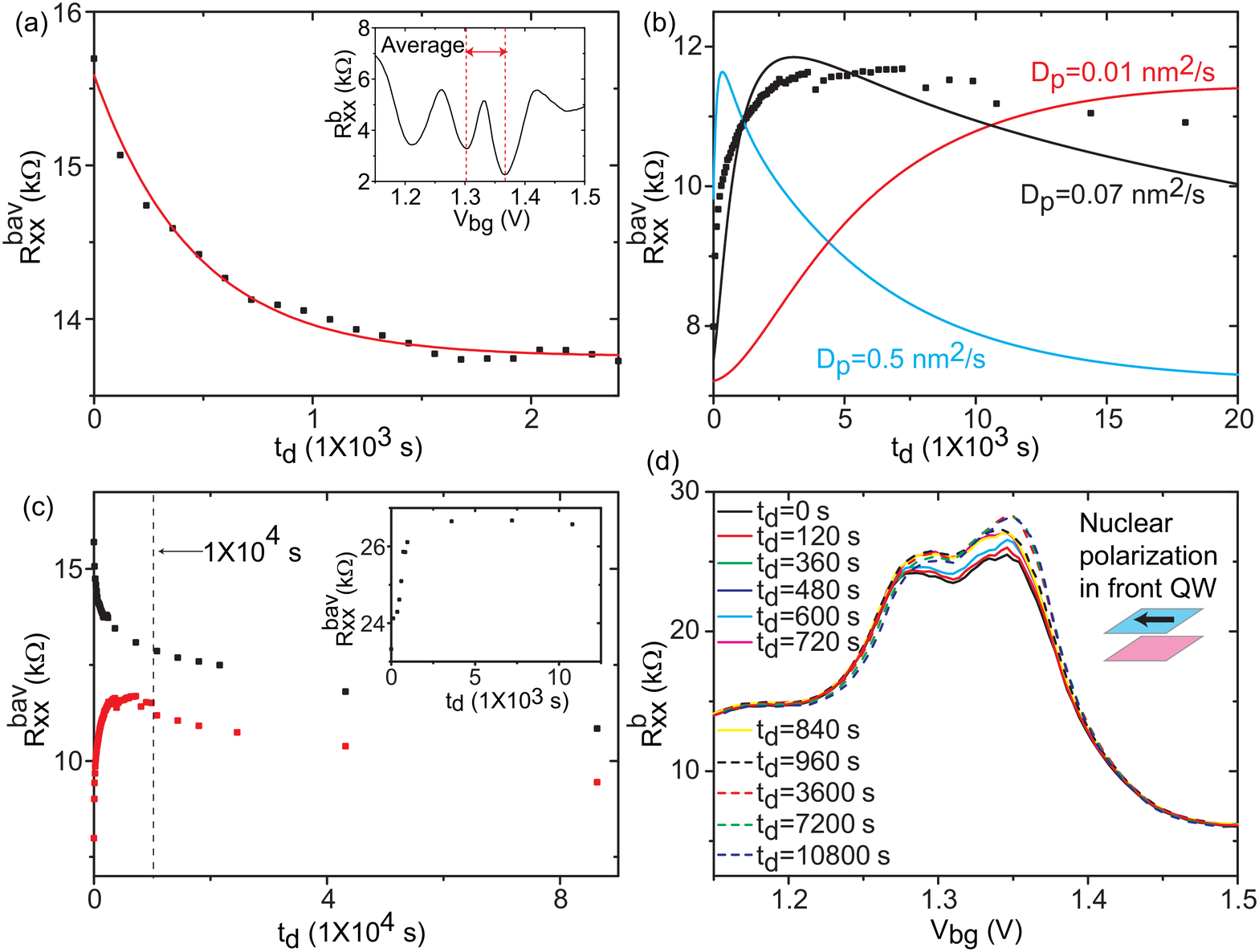}
\caption{(a) Average value of $R_{xx}^{b}$, $ R_{xx}^{bav}$, as a function of $t_d$ after nuclear polarization of the back QW. Inset: the region in which $R_{xx}^{b}$ is averaged, i.e., between the two red dashed lines. The curve is the same as that for $t_p=0$ s in Fig.\ref{figure2} (c). (b) $ R_{xx}^{bav}$ as a function of $t_d$ after nuclear polarization of the front QW. (c) $ R_{xx}^{bav}$ as a function of long $t_d$ after nuclear polarization of the back (black squares) and front QWs (red squares). Inset: $ R_{xx}^{bav}$ as a function of $t_d$ after nuclear polarization of the front QW for $T_p=3600$s. (d) $R_{xx}^{b}$ as a function of $V_{bg}$ for the various values of $t_d$ after nuclear polarization of the front QW for $T_p=3600$s.} 
\label{figure6}
\end{figure}

Average values of $R_{xx}^{b}$, $R_{xx}^{bav}$, are shown in Figs.\ref{figure6}(a) and (b) as functions of $t_d$ with black squares for nuclear-spin polarization in the back and front QWs, respectively. The value of $R_{xx}^{bav}$ averaged in the SPT peak region shown in the inset of Fig.\ref{figure6}(a) is plotted here.

The diffusion coefficient is calculated from the time evolution of $ R_{xx}^{bav}$ in Figs.\ref{figure6}(a) and (b). For the polarization QW, because the DNP polarized nuclear spins distribute inhomogeneously in the in-plane direction, the nuclear-spin diffusion is expected to occur not only in the perpendicular but also in the in-plane directions. The nuclear-spin diffusion in the perpendicular direction is much slower than that in the in-plane direction because of the heterostructure as discussed later and expected in previous reports \cite{Malinowski,Nikolaenko}. Consequently, the time evolution of $ R_{xx}^{bav}$ is practically caused by the nuclear-spin diffusion in the in-plane direction. A relaxation time $T_{R}\approx 460$s is obtained by fitting the data in Fig.\ref{figure6}(a) with an exponential relaxation function (the red solid line in Fig.\ref{figure6}(a)). Although the nuclear-spin distribution is inhomogeneous in the whole Hall bar system, it is assumed that the nuclear-spin distribution becomes homogeneous over the domain-wall width, $\sim 4l$ \cite{Shibata}, after $T_{R}$ \cite{domainwall} because the DNP occurs around the domain walls. Here $l$ is magnetic length. Accordingly, the diffusion coefficient in the in-plane direction, $D_{i}$, is estimated to be $(4l)^2/T_{R}\approx 4{\rm nm^2/s}$ by using $l\approx10$ nm at $B=5.7$T. This $D_{i}$ value is consistent with the previously reported value for bulk GaAs \cite{Paget}. 

For the detection QW, $R_{xx}^{bav}$ is influenced by the nuclear-spin diffusion in the perpendicular direction for $T_d\gg 460$s because the nuclear- spin distribution is almost homogeneous around the domain walls due to the nuclear-spin diffusion in the in-plane direction \cite{inhomogeneous}. The time evolution of $R_{xx}^{bav}$ is therefore calculated as follows. Because nuclear spins are polarized by the contact hyperfine interaction, it is assumed that the distribution of the current-polarized nuclear spins in the front QW, $p(z)$, is proportional to the square of the absolute value of the electron wave function confined in the front QW, $\psi^{f}(z)$:
\begin{eqnarray}
p(z)=\alpha |\psi^{f}(z)|^2,  
\end{eqnarray}
where $\alpha$ is a constant, indicating nuclear polarization rate \cite{Ren}. $z$ is the vertical axis to the QWs. $p(z)$ is used to obtain the time evolution of the distribution of the nuclear polarization at $t$, $P(z,t)$ as 
\begin{eqnarray}
\frac{\partial P(z,t)}{\partial t}=D_{p}\frac{\partial^2 P(z,t)}{\partial z^2}+p(z)(1-\theta(t)) \: \: (t>-T_p), 
\end{eqnarray}
where $D_{p}$ is the diffusion coefficient in the perpendicular direction. The second term of the right side of the equation corresponds to the current-induced DNP per unit time, and $\theta(t)$ is a step function. The spin-lattice relaxation term is neglected because the spin-lattice relaxation time is extremely long, as shown in Figs.\ref{figure4}(d) and (e). This approximation will be justified later. 
It is assumed that the variation of $R_{xx}^{bav}$ is proportional to the integral of the product of the square of the absolute value of the electron wave function confined in the back QW, $\psi^{b}(z)$, and $P(z,t)$ because the resistive detection of nuclear spins is also based on the contact hyperfine interaction \cite{YH}. Consequently, $R_{xx}^{bav}(t)$ is calculated as 
\begin{eqnarray}
R_{xx}^{bav}(t) =A \int_{\rm in \: back \: QW} |\psi^{b}(z) |^2 P(z,t) dz+ R_{xx}^{bav0}. 
\label{deltarxx}
\end{eqnarray}
where A is a constant determined by $\alpha$ to fit experimental data, and $R_{xx}^{bav0}$ is the background value of $R_{xx}^{b}$ \cite{Nakajima2}. 

$R_{xx}^{bav}$ calculated with $D_{p}=0.01$, 0.07, and 0.5${\rm nm^2/s}$ is shown in Fig.\ref{figure6}(b). $R_{xx}^{bav}$ calculated with $D_{p}=0.07{\rm nm^2/s}$ is closer to the measured $R_{xx}^{bav}$ than the other calculated lines. This value of $D_{p}$ is two orders of magnitude smaller than that shown in Fig.\ref{figure6} (a). This small value is similar to that reported by Nikolaenko {\it et al.} for the GaAs/Al$_x$Ga$_{1-x}$As interface of quantum dots \cite{Nikolaenko,cal}. Moreover, it might be explained by the fact that mutual spin flips of the same isotopes are affected by the quadrupole couplings produced by the slight lattice mismatch between the GaAs wells and AlGaAs barriers \cite{Barrett}. 

$R_{xx}^{bav}$ after the DNP in the back and front QWs for long $t_d$ is shown by the black and red squares, respectively,  in Fig.\ref{figure6}(c). 
At less than $\sim 10^4$s, the onset point of the longitudinal resistance decrease is delayed in the detection QW. This indicates the direct demonstration of the nuclear spin diffusion from the polarization QW to the detection QW. On the other hand, $R_{xx}^{bav}$  shows a similar relaxation rate for $t_d\gtrsim 10^4$s. 
This decrease in $R_{xx}^{bav}$ is therefore understood in terms of the spin-lattice relaxation for $t_d\gtrsim 10^4$s. By fitting the data in Fig.\ref{figure6}(c) with exponential relaxation functions, the spin-lattice relaxation times, $T_1$, for the polarization and detection QWs are estimated respectively to be approximately $3.4\times 10^4$s and $7.5\times 10^4$s. Consequently, the nuclear-spin relaxation due to nuclear-spin diffusion is much faster than spin lattice relaxation, substantiating the estimation of the nuclear-spin diffusion without the effect of the spin-lattice relaxation.

To investigate the influence of the uniformity of nuclear-spin distribution on nuclear-spin diffusion, $R_{xx}^{b}$ is measured by following the measurement sequence shown in Fig.\ref{figure4}(a) for $T_p=3600$s, at which $R_{xx}^{b}$ is sufficiently saturated. $R_{xx}^{b}$ at various values of $t_d$ for $T_p=3600$s is shown in Fig.\ref{figure6}(d). In a similar manner to that shown in Fig.\ref{figure4}(c), $R_{xx}^{b}$ increases around the SPT as $t_d$ increases. $R_{xx}^{bav}$ for $T_p=3600$s is shown as black squares in the inset of Fig.\ref{figure6}(c). The data were calculated with the same method as that used for calculating the data in Figs.\ref{figure6}(a) and (b). Although the value of $R_{xx}^{bav}$ is different from that shown in Fig\ref{figure6}(b), the time evolution of $R_{xx}^{bav}$ is very similar to that shown in Fig\ref{figure6}(b). 

\section{Discussion}

As discussed above, our results are different from those presented in Ref.\cite{Nguyen}. In our measurement sequences, the nuclear-spin distribution in the two QWs is initialized after every $R_{xx}^{b}$ measurement, and one-shot measurement is performed to exclude hyperfine interaction \cite{initial}. The careful experimental sequences used in this study results in slow nuclear-spin diffusion in the perpendicular direction. In addition, when $T_p$ is shorter than the time that the longitudinal resistances in the QW are saturated, it is possible to observe nuclear-spin diffusion in the in-plane direction. Consequently, an inhomogeneous nuclear-spin distribution makes it possible to measure in-plane and perpendicular nuclear-spin diffusions simultaneously by using double QWs.

Finally, Domain structures and nuclear-spin diffusion are discussed as follows. The domain size is reportedly $0.5\sim3\mu$m\cite{Hayakawa,Verdene,Dini}, which is larger than the diffusion length in this experiment, $\sqrt{D_{i}T_1} \approx400$nm. The distribution of the nuclear polarization in the whole Hall bar is therefore inhomogeneous, even at $t_d>T_1\approx10^4$s. It is noteworthy that the shapes of $R_{xx}^{b}$ of the polarization and detection QWs around the SPT in Figs.\ref{figure2} and \ref{figure4} differ. The distribution of the nuclear polarization in the polarization (back) QW depends mainly on the DNP around the domain walls in the back QW and the nuclear-spin diffusion in the in-plane direction. In contrast, the distribution of the nuclear polarization in the detection (back) QW is affected primarily by the DNP around the domain walls in the polarization (front) QW and the nuclear- spin diffusion in the in-plane and perpendicular directions. The difference in the shape of $R_{xx}^{b}$ therefore results from the difference in the distributions of nuclear polarization in the in-plane direction in the polarization and detection QWs. Furthermore, the shape of $R_{xx}^{b}$ in Fig.\ref{figure6}(d) is more similar to that in Figs.\ref{figure4}(b) and (d) than to that in Figs.\ref{figure4}(c) and (e). This may indicate the existence of a spatial steady state of nuclear-spin distribution independently of the polarization caused by the nuclear-spin diffusion or the DNP. 

\section{Summary}

In summary, nuclear-spin diffusion in a double QW was investigated by using resistive detection at $\nu=2/3$ SPT. From the different time evolutions of longitudinal resistance in polarization and detection QWs, anisotropic nuclear-spin diffusion was clarified. The knowledge of this anisotropy is important to implement nanoscale devices, particularly the future quantum computation based on nuclear spins and electron spins.

\begin{acknowledgments}

We greatly appreciate K. Muraki and NTT for providing high-quality wafers and fruitful discussions, S. Ichinokura for the sample fabrication, and A. Sawada, M.-H. Nguyen, S. Tsuda, K. Yamaguchi, and H. M. Fauzi for their fruitful discussions. 
Part of this work is supported by the Grant-in-Aid for Scientific Research C (No. 26400307) from the Japan Society for the Promotion of Science (JSPS).

\end{acknowledgments}

\appendix

\section{Multi-measurement of Time Evolution of Longitudinal Resistance}

\begin{figure}[hb]
\begin{center}
\includegraphics[width=1\columnwidth]{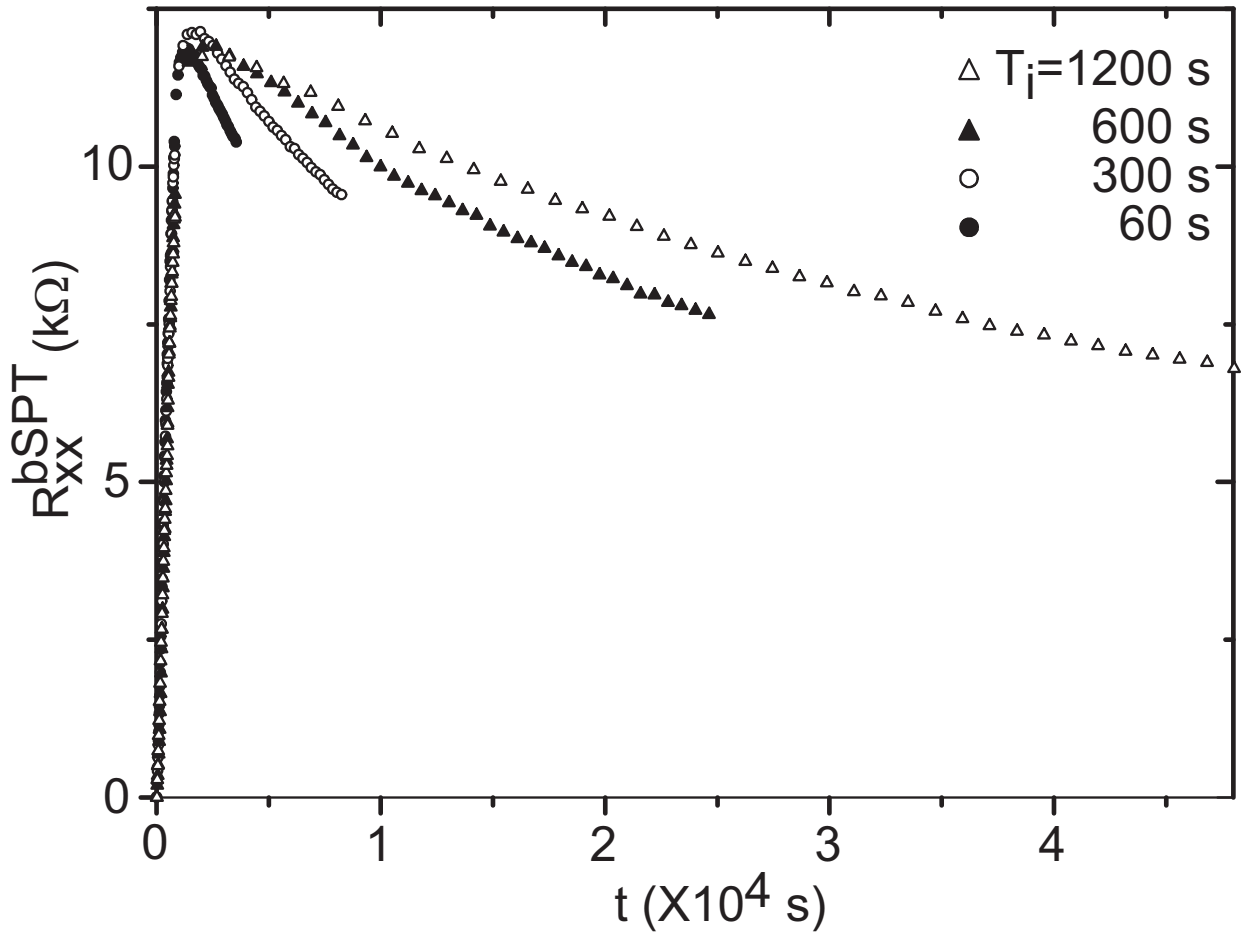}
\end{center}
\caption{
Time evolution of the longitudinal resistance at $\nu=2/3$ spin phase transition (SPT) in the back QW, $R_{xx}^{bSPT}$, measured for various values of interval times $T_i=$60, 180, 600, and 1200 s. Current-induced dynamic nuclear polarization is performed in the front QW under the depletion of the back QW.
} 
\label{olddata}
\end{figure}

A small bias current is applied to a QW at the SPT to measure the variation of the longitudinal resistance and detect nuclear spin polarization. Because the longitudinal resistance is almost constant when a small current is applied, it was previously reported that a small current does not strongly polarize nuclear spins \cite{Nguyen}. Accordingly, in the present study, the influence of multiple measurements on the longitudinal resistance and the nuclear polarization was investigated.

The sequence for measuring the time evolution of the longitudinal resistance at $\nu=2/3$ SPT in the back QW is described as follows. First, the nuclear spins in the front and back QWs are initialized alternately; at $\nu=0.9$ and $I=50$ nA applied to the front (back) QW for 600 s, nuclear spins in the front (back) QWs are depolarized because of the Goldstone mode of the Skyrmions, and then the back (front) QW is depleted at $V_{bg}=-0.8$ V ($V_{fg}=-1.2$ V), similarly to the measurement sequences described in the main text. Subsequently, starting from $t=0$ s, the nuclear spins in the front QW are polarized for 400 s by application of $I=50$ nA at the $\nu=$2/3 SPT (polarization), and the back QW is depleted during polarization. The longitudinal resistance of the front (polarization) and back (detection) QWs at the $\nu=$2/3 SPT is then measured at a regular interval time of $10$ s during polarization, by application of $I=1$ nA for 5 s. Finally, both QWs are depleted. The longitudinal resistance of the polarization and detection QWs at the $\nu=$2/3 SPT with various values of a regular interval time $T_i$ are then measured by application of $I=1$ nA for 5 s. The time evolution of the longitudinal resistance of detection QW at the $\nu=$2/3 SPT,  $R_{xx}^{bSPT}$, at $T_i=60$, 180, 600 and 1200 s is shown in Fig.~\ref{olddata}. $ R_{xx}^{bSPT}$ increases continuously at all values of $T_i$ after the nuclear polarization is terminated. This is the same result as that stated in the main text and in a previous report \cite{Nguyen}. However, when the value of $T_i$ becomes large, the decrease rate of $ R_{xx}^{bSPT}$ reduces. Therefore, $R_{xx}^{bSPT}$ keeps a high value for a longer $T_i$. This result indicates that, although the influence of the measurement on the nuclear spin distribution is negligible for one-shot measurement of longitudinal resistance at a small current, the nuclear spin distribution is affected significantly by multi-measurements of longitudinal resistance.



\begin{thebibliography}{99}
\bibitem{Hirayama}
For spatially homogeneous nuclear polarization, we expect a shift of the SPT peak as confirmed experimentally at extremely low temperatures \cite{Fauzi} and with DNP induced by optical irradiation \cite{Akiba1,Akiba2}. In contrast to these results, the SPT peaks are always broad in the case of current-induced DNP at $\nu =2/3$ SPT, clearly suggesting spatially inhomogeneous nuclear polarization.
\bibitem{SAS}
The energy separation between the symmetric and antisymmetric states, $\Delta_{SAS}$, at the total electron density in the double QWs, $1.45\times10^{15}\rm{m^{-2}}$ is estimated to be 1.5meV.
\bibitem{initial}
In our measurements, the nuclear-spin distribution in the two QWs is always initialized after every $R_{xx}^{b}$ measurement. Although the hyperfine interaction between the nuclear spins and electron spins is negligible at a small current for one-shot measurement, it modifies the experimentally obtained results for multi-measurement sequence (see the Appendix A).
\bibitem{tp}
When a large current is applied at the SPT, $R_{xx}^{b}$ increases. $R_{xx}^{b}$ saturates after a current-application time of $\sim 600$s, suggesting homogeneous nuclear polarization around the domain walls is established within this time scale. On the other hand, the nuclear-spin distribution is inhomogeneous around domain walls at $T_p=300$s. Owing to this spatial inhomogeneity, the nuclear-spin diffusion occurs not only in the perpendicular direction but also in the in-plane direction.
\bibitem{replace}
Similar experimental results were obtained by replacing the back QW with the front QW. However, the larger signal obtained in the case of the back-QW detection is probably attributable to the difference in the domain structures of the two QWs. $R_{xx}^{b}$ is therefore measured in this study.
\bibitem{region}
We calculated $R_{xx}^{bav}$ for different averaged regions, and we confirmed that the time evolution was the same feature independently of the selection of the averaged region.
\bibitem{domainwall}
The nuclear spins are polarized around the domain walls by the current flow in them (edges of the domains), and $R_{xx}^{b}$ increases. When the polarized nuclear-spin distribution becomes homogeneous around the domain walls, $R_{xx}^{b}$ is affected only slightly by the in-plane nuclear diffusion.
\bibitem{inhomogeneous}
In the whole Hall bar, the distribution of the nuclear polarization is inhomogeneous, even at the large value of $t_d$ (details of which are discussed later).
\bibitem{YH}
In Ref.\cite{Yusa,Kumada}, the experimental result can substantially be explained by the simple assumption that the variation of the longitudinal resistance is proportional to the degree of nuclear spin polarization.
\bibitem{cal}
It is necessary to solve the two-dimensional diffusion equation for nuclear-spin distribution to obtain precise values of $D_{i}$, and $D_{p}$.
\end{thebibliography}
\end{document}